\documentstyle{article} 	
\begin{document}

\title{{\small\centerline{March 1994 \hfill DOE-ER\,40757-045 / CPP-94-13}}
\medskip
{\bf What is the Equivalence theorem really?}}

\author{\bf Palash B. Pal\\
\normalsize \em Center for Particle Physics, University of Texas,
              Austin, TX 78712, USA}

\date{}
\maketitle

\begin{abstract} \normalsize\noindent
The precise statement of the equivalence theorem, between the
longitudinally polarized states of a massive gauge boson and the
corresponding goldstone mode, is discussed when the amplitude in
question depends on masses of other particles (e.g., the Higgs boson)
in the theory.
\end{abstract}
\bigskip\bigskip

\section{Introduction}
In spontaneously broken gauge thoeries, some gauge bosons acquire
mass. As a consequence,  a longitudinal degree of polarization appears
for each of them, which is accompanied by the disappearance of a
spin-0 boson from the spectrum of the unbroken Lagrangian. Thus, in a
vague sense, this spin-0 goldstone mode is ``eaten up'' by the gauge
boson and it manifests itself as the longitudinal polarization state
in the broken phase. One therefore vaguely expects that the amplitude
of any process involving the longitudinal vector boson is equivalent
to the amplitude of the same process calculated with the goldstone
mode replaced in the outer lines for the longitudinal vector boson
lines. This statement is formalized into the {\em Equivalence
theorem} \cite{CLT74,LQT77,ChGa85,Wil88,BaSc90,Vel90,DuNa94}.
There exist various precise statements of this theorem.
The purpose of this article is to discuss them and suggest one which
covers all known cases and follows from simple physical arguments.

Although the equivalence theorem should be valid for gauge models in
general, we use the standard electroweak model here for the sake of
notational simplicity. Here, $W^\pm$ and $Z$ are the massive gauge
bosons, and let us denote by $w^\pm$ and $z$ the Higgs-Goldstone modes
eaten up by them. We can write for any amplitude
	\begin{eqnarray}
{\cal A} (W^\pm_\parallel, Z_\parallel, \cdots) =
{\cal A} (w^\pm, z, \cdots) + {\cal A}'  \,,
\label{A'}
	\end{eqnarray}
where the subscript on the gauge bosons imply their longitudinal
polarization states, and
the ellipsis denote any other particle in the model, which include the
Higgs boson and the fermions. If this equation is taken to be the
definition of the quantity ${\cal A}'$, equivalence theorem is in effect a
statement about its value. Most simply and commonly, one
encounters the following statement about
${\cal A}'$~\cite{LQT77,ChGa85,Wil88,Vel90}:
	\begin{eqnarray}
{\cal A}' = O(M_W/E) \,,
\label{MW/E}
	\end{eqnarray}
where $E$ is the energy of the gauge boson.  This means that, at
energies much larger than $M_W$, one can replace the amplitude
involving $W^\pm_\parallel$ and $Z_\parallel$ by the amplitude
involving $w^\pm$ and $z$.

This statement is sufficient when the gauge boson is the only massive
particle in the amplitude, but is incomplete and useless if the relevant
amplitude depends on the mass of the Higgs boson and the fermions. One
needs to know the behaviour of ${\cal A}'$ as a function of these masses in
order to make use of the theorem. For example, one can ask whether at
a given value of $E$, the quantity ${\cal A}'$ can be neglected for any
value of the Higgs mass $M_H$, or does $M_H$ has to be in some special
range? Ad hoc answers to such questions have been given by various
authors. In what follows, we give examples to show that all such
prescriptions can be understood in one simple statement of the
equivalence theorem.

First, let us explain why one needs $E \gg M_W$ in order that ${\cal
A}'$ can be neglected.  For any value of
$M_W \neq 0$, the longitudinal components of the $W^\pm$-bosons
are physical states.  The ``Higgs-Goldstone'' modes $w^\pm$
 are unphysical. On the other hand, for
$M_W=0$,  the situation reverses since the $w^\pm$ are the physical
states but the $W^\pm_\parallel$ are not, as discussed below.
The equivalence theorem then merely states that all observables are
continuous in the limit $M_W\to 0$. In other words, in that limit, the
amplitudes for any process with $W^\pm_\parallel$
bosons is the same (apart from a phase maybe) with the amplitudes for
the corresponding processes where the $W^\pm_\parallel$ are replaced
by $w^\pm$. The limit $M_W\to 0$, of course, is realized physically if
one deals with energies for which $M_W/E \to 0$.

Now, the $W$-mass is given by
	\begin{eqnarray}
M_W = {1\over 2} gv \,,
	\end{eqnarray}
where $g$ is the SU(2)$_L$ coupling constant and $v$ is the vacuum
expectation value (VEV) of the Higgs field. Thus, the mathematical
limit of vanishing $W$-mass can be obtained in two ways:
	\begin{eqnarray}
M_W \to 0 \quad \mbox{if} \left\{ \begin{array}{l} \mbox{either $g \to
0$} \\ \mbox{or $v \to 0$.}
\end{array} \right.
\label{2lims}
	\end{eqnarray}
Bagger and Schmidt \cite{BaSc90} clearly emphasized that the
equivalence theorem has different interpretation in these two limits.
For $g=0$, we have only a global symmetry, so the Higgs mechanism
should not work and the $w^\pm$ and $z$ should be physical states. On
the other hand, for $v=0$, we have a gauge theory whose symmetry is
not broken, and this is why the $w^\pm$ and $z$ appear as physical
states.

We will show that, because of these two limits,  we should expect {\em
two equivalence theorems\/} instead of {\em the equivalence theorem}, viz.,
	\begin{eqnarray}
\lim_{g \to 0} {\cal A}' = 0 \,,
\label{thm1}\\
\lim_{v \to 0} {\cal A}' = 0 \,.
\label{thm2}
	\end{eqnarray}
We will see that each of these cases yields a statement of the
equivalence theorem, and in the presence of Higgs and fermion masses
the two statements are different in the sense that they imply the
equality of different parts of the amplitudes
${\cal A}(W^\pm_\parallel,Z_\parallel,\cdots)$ and
${\cal A}(w^\pm,z,\cdots)$. Moreover, whereas different ad-hoc
prescriptions have been suggested to modify the statement of
equivalence theorem as given in Eqs.\ (\ref{A'}) and (\ref{MW/E}), we
will show that all these statements can be derived from the two limits
given in Eqs.\ (\ref{thm1}) and (\ref{thm2}).

\section{Examples of various processes}\label{s:examples}
In this section, we consider the amplitudes of various $2\to2$
scattering processes
involving longitudinal gauge bosons and of the corresponding proceses
involving the Higgs-Goldstone modes. In particular, we discuss
amplitudes which depend, in addition to the gauge boson masses, on the
Higgs boson mass only, and show in what sense the
equivalence theorem is vindicated in these calculations.

\subsection{$W^+_\parallel W^-_\parallel \to HH$.}
Consider, for example, the process $W^+_\parallel W^-_\parallel \to
HH$. Calculation of the
amplitude exists in the literature \cite{DKW88,Kal88}. In terms of the
usual Mandelstam variables $s$, $t$ and $u$, one obtains
	\begin{eqnarray}
{\cal A} (W^+_\parallel W^-_\parallel \to HH) &=& {g^2 \over 4 (1 -
4M_W^2/s)} \left\{ {M_H^2 \over M_W^2} \left[ 1 + {3M_H^2 \over s -
M_H^2} + M_H^2 \left( {1 \over t- M_W^2} + {1 \over u - M_W^2}
\right) \right] \right. \nonumber \\*
&&  + 2 \left[ 1 - {9M_H^2 \over s-M_H^2} + 4 {M_W^2 \over s} \left(
1 + {3M_H^2 \over s - M_H^2} \right) \right] \nonumber \\*
&& \left. + 2 \left[ s - 2M_H^2 - 4M_W^2 + 8 M_W^4/s \right]
\left( {1 \over t- M_W^2} + {1 \over u - M_W^2}
\right) \right\} \,,
\label{WW->HH}
	\end{eqnarray}
whereas~\cite{DKW88}
	\begin{eqnarray}
{\cal A} (w^+ w^- \to HH) &=& {g^2 \over 4}
\left\{ {M_H^2 \over M_W^2} \left[ 1 + {3M_H^2 \over s -
M_H^2} + M_H^2 \left( {1 \over t- M_W^2} + {1 \over u - M_W^2}
\right) \right] \right. \nonumber \\*
&& \left. + 2 \left[ 1 + (s - M_H^2)
\left( {1 \over t- M_W^2} + {1 \over u - M_W^2}
\right) \right] \right\} \,,
\label{ww->HH}
	\end{eqnarray}
As the authors of Ref.\ \cite{DKW88} observe, the terms enhanced by
$M_H^2/M_W^2$ agree in the two amplitudes for the limit $M_W^2/s \to
0$. The remaining terms agree only when the extra condition $M_H^2/s
\to 0$ is imposed.

It is easy to understand this in terms of the two limits mentioned in
Eq.\ (\ref{2lims}). Remember that the Higgs mass is given by
	\begin{eqnarray}
M_H^2 = 2 \lambda v^2
\label{MH}
	\end{eqnarray}
in terms of the Higgs quartic coupling $\lambda$. Thus,
	\begin{eqnarray}
{M_H^2 \over M_W^2} = {8 \lambda \over g^2} \,.
\label{MH2/MW2}
	\end{eqnarray}
Because of the $1/g^2$ factor here, the terms involving the factor
$M_H^2/M_W^2$ in Eqs.\ (\ref{WW->HH}) and (\ref{ww->HH}) are really {\em
independent\/} of $g$. Thus, these are the only terms which survive in
the amplitude if we take the limit $M_W\to 0$ by letting $g$ going to
zero, and therefore have to be equal.

On the other hand, if we take $v\to 0$, we need not consider only the
$g$-independent terms. However, Eq.\ (\ref{MH}) tells us that in this
limit, $M_H$ also goes to zero, so equivalence theorem concerns only
the terms with ${M_H/ \sqrt s} \to 0$ in addition to ${M_W/\sqrt s}
\to 0$.

\subsection{$W^+_\parallel W^+_\parallel$ scattering.}
These amplitudes have been discussed in Ref.\ \cite{BCHP90}. In this
case, one obtains
	\begin{eqnarray}
{\cal A} (w^+w^+ \to w^+w^+) &=& {g^2 \over 4} \left\{ - {M_H^2 \over
M_W^2} \left( {t \over t-M_H^2} + {u \over u-M_H^2} \right) \right.
\nonumber \\*
&& + \left. 4 \sin^2 \theta_W \left( {s-u \over t} + {s-t \over u} \right) +
{\cos^2 2\theta_W \over \cos^2 \theta_W} \left( {s-u \over t-M_Z^2} +
{s-t \over u-M_Z^2} \right) \right\} \,,
\label{wwscatt}
	\end{eqnarray}
whereas the amplitude for the scattering using the longitudinal gauge
bosons has the following extra terms for $M_W^2\ll s$,
	\begin{eqnarray}
{\cal A}' = - g^2 M_H^2
\left\{ {4 \over s} + \left( 1 + {M_W^2 \over M_H^2} + {2M_H^2 \over
s} \right) \left( {1 \over t-M_H^2} + {1
\over u-M_H^2} \right) \right\} \,.
\label{wwA'}
	\end{eqnarray}

Once again, notice that the leading terms in Eq.\ (\ref{wwscatt}) are
proportional to $g^2M_H^2/M_W^2$, which means that they do not go to
zero as $g\to 0$ because of the mass relation in Eq.\ (\ref{MH2/MW2}).
The terms in ${\cal A}'$, on the other hand, all go to zero in this limit,
so that equivalence theorem is verified. Note that the terms in Eq.\
(\ref{wwscatt}) involving $\theta_W$ are irrelevant for this limit, because
they vanish for $g\to 0$.

On the other hand, if we take $v\to 0$, it implies $M_H\to 0$
alongwith $M_W\to 0$, as mentioned earlier. In this case, equivalence
theorem implies equality of terms having higher powers of the gauge
coupling constant as well. As we see from the expression of ${\cal A}'$ in
Eq.\ (\ref{wwA'}), this is indeed true since ${\cal A}'$ vanishes
altogether for $M_H\to 0$.

In passing, we also want to note \cite{BCHP90}, using Eqs.\
(\ref{wwscatt}) and (\ref{wwA'}), that ${\cal A}'/{\cal A}$ goes to
zero also for $M_H^2 \to \infty$, so that ${\cal A}'$ can be neglected
in this case as well. The implication of this will be discussed in
Sec.~\ref{s:decoup}.

\subsection{$W^+_\parallel W^-_\parallel \to Z_\parallel Z_\parallel$.}
The amplitude for $W^+_\parallel W^-_\parallel \to Z_\parallel
Z_\parallel$ has been calculated in Refs.\ \cite{BeLl87,DuNa94}.
Putting $M_W,M_Z=0$ as is required to verify the equivalence theorem,
we obtain from these calculations
	\begin{eqnarray}
{\cal A} (W^+_\parallel W^-_\parallel \to Z_\parallel
Z_\parallel) &=& {g^2 \over 4 tu (s-M_H^2)}
\times \left[ \left( {M_H^2 \over M_W^2} + 6 \right) s^3
+ \left( - 4  + 6 {M_Z^2 \over M_W^2} \right) M_H^2 s^2 \right.
\nonumber\\*
&& \left. + \cos^2\theta \left\{ \left( {M_H^2 \over M_W^2} + 2
\right) s^3 - \left( 8 + 6 {M_Z^2 \over M_W^2} \right) M_H^2 s^2
\right\}  \right] \,,
\label{WWZZ}
	\end{eqnarray}
where $\theta$ is the scattering angle in the center-of-mass frame.
The amplitude calculated using the Higgs-Goldstone modes in the outer
lines, evaluated for $M_W=M_Z=0$, is given by~\cite{DuNa94,DuNa94b}
	\begin{eqnarray}
{\cal A} (w^+ w^- \to zz ) &=& {g^2 \over 4 tu (s-M_H^2)}
\times \left[ \left( {M_H^2 \over M_W^2} + 6 \right) s^3
+ \left( - 6  + 4 {M_Z^2 \over M_W^2} \right) M_H^2 s^2 \right.
\nonumber\\*
&& \left. + \cos^2\theta \left\{ \left( {M_H^2 \over M_W^2} + 2
\right) s^3 - \left( 6 + 4 {M_Z^2 \over M_W^2} \right) M_H^2 s^2
\right\}  \right] \,.
\label{wwzz}
	\end{eqnarray}
Thus~\cite{DuNa94}
	\begin{eqnarray}
{\cal A}' = {g^2\over 2} \left( 1 + {M_Z^2 \over M_W^2} \right) {M_H^2
\over s - M_H^2} \,.
\label{wwzzA'}
	\end{eqnarray}
Once again, it is easy to see that ${\cal A}'$ vanishes either when $g=0$,
in which case only the terms involving $M_H^2/M_W^2$ are important in
the amplitudes of Eqs.\ (\ref{WWZZ}) and (\ref{wwzz}); or when $v=0$,
which would imply $M_H=0$ in addition to $M_W=M_Z=0$. Once again, note
that for $M_H\to \infty$, ${\cal A}'$ is negligible in comparison with the
amplitude calculated in either way. This is the subject of
Sec.~\ref{s:decoup}.

\section{The case of $M_H^2 \to \infty$.}\label{s:decoup}
In the calculations of some amplitudes which depend on $M_H$, the
authors have noticed that the difference between ${\cal A}
(W^\pm_\parallel, Z_\parallel, \cdots)$ and ${\cal A} (w^\pm, z, \cdots)$
become negligibly small even when $M_H^2/s\to \infty$
\cite{BCHP90,DuNa94,DuNa94b}.
This property has also been assumed to be true in demonstrating the
equivalence theorem \cite{Wil88,Vel90} in some cases. In
this section, we show which
class of amplitudes show this property and what is the relation of
this result with the equivalence theorem. Our conclusion is that this
property is not, in general, a consequence of the equivalence theorem,
but rather follows from an application of the decoupling
theorem~\cite{ApCa75}.

In a technical sense, the decoupling theorem is not expected to work
for $M_H \to \infty$ since the Higgs boson mass comes from physics at
the electroweak scale. In other words, taking $M_H \to \infty$
implies, via Eq.\ (\ref{MH}), $\lambda \to \infty$, which
implies that the Higgs field has very large self-coupling. This
invalidates perturbative calculations on which the above analysis is
based.

However, one can blindly take a Feynman diagram, calculate its
amplitude perturbatively, consider it as a mathematical expression and
ask what happens to it if we take the mathematical limit of $\lambda
\to \infty$. If this limit is zero, we can say that the particular
diagram {\em decouples\/} in the limit of $M_H \to \infty$, although we
emphasize that the physical significance of this statement is not
clear.
Now, if we deal with a process where all diagrams which depend on
$M_H$ decouple in the limit $M_H \to \infty$, then taking the limit we
will obtain an expression which depends only on the masses of the
gauge bosons. The equivalence theorem should work with this
expression when we take the limit of vanishing gauge boson masses.

We can see examples of this statement in the processes described in
Sec.~\ref{s:examples}. For $W^+_\parallel W^+_\parallel$ scattering as
well as $W^+_\parallel W^-_\parallel \to Z_\parallel Z_\parallel$,
there are diagrams with a 4-point interaction or with $W$-exchange in the
$t$- and $u$-channels which are independent of $\lambda$. The only
$\lambda$ dependence comes from the diagram with $s$-channel Higgs
exchange, which goes like $1/M_H^2$ for large Higgs mass, which means
$1/\lambda$. Thus, this diagram decouples for large Higgs mass. The
equivalence theorem should work for the other diagrams in this limit,
whose amplitudes depend only on the gauge boson masses.
Of course, this is easily seen from
the expression for ${\cal A}'$ in Eq.\ (\ref{wwA'}) as well as from Eq.\
(\ref{wwzzA'}), and this was even noted by the authors who calculated
these processes \cite{BCHP90,DuNa94}.

On the contrary, if we consider the process $W^+_\parallel
W^-_\parallel \to HH$, the $s$-channel Higgs exchange diagram now has
an extra power of $\lambda$ coming from the cubic vertex of Higgs
bosons. This diagram is a constant for $\lambda \to \infty$, i.e., it
does not decouple. In this case, one should not expect the equivalence
thoerem to hold for large Higgs mass, and in fact a scrutiny of Eqs.\
(\ref{WW->HH}) and (\ref{ww->HH}) shows that it doesn't.

\section{Zeroes of the amplitude: the case of $\gamma\gamma \to
W^+_\parallel W^-_\parallel$}\label{s:zeroes}
In this section, we give an example where the equivalence theorem may
appear to be invalidated due to some cancellations in the amplitude.
Consider the scattering process $\gamma(k) \gamma (k') \to
W^+_\parallel (p) W^-_\parallel (p')$. In the center of energy frame,
we can write the various momenta as
	\begin{eqnarray}
	\begin{array}{rclrcl}
k^\mu &=& E (1,0,0,1) \,, \quad\quad & k'^\mu &=& E (1,0,0,-1) \,, \\
p^\mu &=& E (1,\beta \sin \theta,0,\beta \cos \theta) \,, &
p'^\mu &=& E (1,-\beta \sin \theta,0,-\beta \cos \theta) \,,
	\end{array}
\label{kk'pp'}
	\end{eqnarray}
where $\beta$ is the magnitude of the 3-velocity of the $W$-bosons.
The polarization vectors are taken to be
	\begin{eqnarray}
	\begin{array}{rclrcl}
\varepsilon^\mu &=& (0,\sin \varphi, e^{i\delta} \cos \varphi, 0) \,,
\quad\quad &
\varepsilon'^\mu &=&  (0,\sin \varphi', e^{i\delta'} \cos \varphi', 0)
\,, \\
\epsilon^\mu &=& {E \over M_W} (\beta, \sin \theta,0,\cos \theta) \,,
&
\epsilon'^\mu &=& {E \over M_W} (\beta, -\sin \theta,0,-\cos \theta) \,.
	\end{array}
	\end{eqnarray}
Note that $\varepsilon^\mu$ and $\varepsilon'^\mu$ represent the most
general choice of the polarization
vectors subject to the electromagnetic gauge invariance. Previous
calculations of this amplitude have used some very special
polarization states \cite{BoJi92,Dic**}. As we will show, our general
choice increases our understanding of the equivalence theorem.

The tree-level diagrams for the process include a 4-point interaction,
and $W$-exhange graphs in the $t$ and the $u$ channels.
A simple calculation yields for the amplitude of the process
	\begin{eqnarray}
{\cal A} (\gamma\gamma \to W^+_\parallel W^-_\parallel) = - {2e^2 \over 1 -
\beta^2 \cos^2 \theta} &\times& \left[ \left( \sin \varphi \sin \varphi' +
e^{i(\delta + \delta')} \cos \varphi \cos \varphi' \right) \left\{ -1
+ (2 - \beta^2) \cos^2 \theta \right\} \right. \nonumber \\*
&& \left. \quad + 2 (2 - \beta^2) \sin \varphi \sin \varphi' \sin^2
\theta
\vphantom{e^{i(\delta + \delta')}} \right] \,.
	\end{eqnarray}
On the other hand, the amplitude using the Higgs-Goldstone modes are
obtained to be
	\begin{eqnarray}
{\cal A} (\gamma\gamma \to w^+ w^-) = {2e^2 \over 1 -
\beta^2 \cos^2 \theta} &\times& \left[ \left( \sin \varphi \sin
\varphi' + e^{i(\delta + \delta')}
\cos \varphi \cos \varphi' \right)
\left\{ - {\scriptstyle 1\over\scriptstyle 2}(1+ \beta^2)
+  \beta^2 \cos^2 \theta \right\} \right. \nonumber \\*
&& \left. \quad + 2 \beta^2 \sin \varphi \sin \varphi' \sin^2 \theta
\vphantom{e^{i(\delta + \delta')}} \right] \,.
\label{gg->ww}
	\end{eqnarray}
Notice that the two amplitudes are indeed equal in the limit $M_W\to
0$, i.e., $\beta \to 1$ for general values of the polarization vectors
for the photons, and thus equivalence theorem is verified.

There is, however, one special case. This is when the two photons have
the same circular polarization. This corresponds to the choice of
$\delta=\delta'=\pi/2$, $\varphi=-\varphi'=\pi/4$. Notice that in this
case, the exact amplitude becomes
	\begin{eqnarray}
{\cal A} (\gamma\gamma \to W^+_\parallel W^-_\parallel) =
2e^2 \times \left[ {1-\beta^2  \over 1 - \beta^2 \cos^2 \theta}
\right] \,.
\label{amplcirc}
	\end{eqnarray}
The amplitude of Eq.\ (\ref{gg->ww}) is only half as large. Thus,
looking at these amplitudes, one may want to conclude that the equivalence
theorem does not hold in this case~\cite{BoJi92}.

One can argue about this statement. Certainly it is true that Eqs.\
(\ref{thm1}) and (\ref{thm2}) are valid still, because ${\cal A}'$ also
contains the factor $1-\beta^2$, and therefore vanishes when $M_W=0$,
i.e., $\beta=1$. So, in this sense, the equivalence theorem is still
valid. What is different from the general case is that, here
not only ${\cal A}'$ vanishes in the limit, but so do the amplitudes
themselves. Thus,  in some sense here the equivalence theorem has no
content, in that one cannot simply calculate the amplitude of
$\gamma\gamma \to w^+ w^-$ and use it for the leading term of
the amplitude of $\gamma\gamma \to W^+_\parallel W^-_\parallel$. Such
a situation arises even for the most general photon polarization
vector if one wants to calculate the process $\gamma Z_\parallel \to
W^+_\parallel W^-_\parallel$, which vanishes for $M_W=M_Z=0$, and
therefore the equivalence theorem is of no useful
consequence.~\footnote{I am grateful to D.~A.
Dicus for checking this statement by explicit calculation.}

There is another important point regarding the expresssion of Eq.\
(\ref{amplcirc}). Suppose someone is interested
in calculating the forward scattering amplitude of the process only
using circularly polarized photons. In this case, one would put
$\theta=0$ from the very beginning.  Looking back at Eq.\
(\ref{amplcirc}), we now notice that the term in the square bracket is
unity, so that the amplitude will turn out to be $2e^2$. On the other
hand, using the Higgs-Goldstone modes, one would obtain an amplitude
which is half as large.  For this special case, one might then conclude
that the equivalence theorem is wrong.

The point to emphasize is that, equivalence theorem is not expected to
work for forward or backward scattering, i.e., when the scattering
angle is $0$ or $\pi$.  We said earlier that the equivalence thoerm
concerns the limit $M_W\to 0$. Physically, this means that $M_W^2$ has
to be much smaller than {\em all} the kinematical variables of the
problem, viz., the invariant products of different external momenta
(for $2\to 2$ scattering problems, these are equivalent to the
Mandelstam variables). In this problem, for example, using Eq.\
(\ref{kk'pp'}), one finds
	\begin{eqnarray}
{M_W^2 \over p \cdot k} = {1 - \beta^2 \over 1- \beta \cos \theta} \,.
	\end{eqnarray}
For the case of forward scattering,  $\theta=0$, this ratio in fact
has a limiting value of 2 when $\beta\to 1$. Thus, the limit $M_W\to
0$ is certainly not realized here. Similarly, for backward scattering,
$\theta=\pi$, it is easily seen that one faces the same problem with
the ratio $M_W^2/p\cdot k'$.  The
lesson learned is this: the equivalence theorem may not work for
some special choice of external momenta. One should be careful to check
that the choice does not imply the smallness of any kinematical
variable. For $2\to2$ scattering, this means that one needs to ensure
$M_W^2 \ll s, |t|, |u|$.

\section{Summing up}
We have thus shown that the equivalence theorem can always be
understood in terms of the limits $g\to 0$ or $v \to 0$, as mentioned
in Eqs.\ (\ref{thm1}) and (\ref{thm2}). These prescriptions, which can
be derived from simple physical arguments \cite{BaSc90} mentioned in
the Introduction, remove the arbitrariness of the various ad-hoc
statements appearing in the literature about the validity of
equivalence theorem in amplitudes dependent on the Higgs boson mass
and fermion masses. Moreover, we have emphasized that these two
limiting procedures really have different physical and mathematical
content. Physically, given the energy, whether or not one can use the
limit $v\to 0$ depends on what the Higgs mass really is. Thus, unless
the Higgs mass is known, we cannot use this version of the equivalence
thorem, i.e., Eq.\ (\ref{thm2}), for a given energy. Mathematically,
the two limits produce different limiting amplitudes, and therefore
the part of the total amplitude equated with the
corresponding amplitude using the Higgs-Goldstone modes is different
for the two limiting cases.

There may be processes
whose amplitude vanishes altogether for $M_W=M_Z=0$, In
Sec.~\ref{s:zeroes}, we have shown that for such processes, although
the equivalence theorem holds, its content is trivial. It merely
asserts that the amplitude with the corresponding Higgs-Goldstone
modes also vanish for $M_W=M_Z=0$, but does not say anything about the
non-vanishing terms which occur for the non-vanishing values of the
gauge boson masses.

Throughout the paper, we have used examples where the Higgs mass is
the only mass other than the gauge boson masses. But from our
discussion, the generalization for amplitudes involving fermion masses
$m_f$ is obvious. For the limit $g\to 0$, the equivalence theorem
dictates that the terms involving $g^2m_f^2/M_W^2$ (which are
really independent of $g$) should be equal irrespective of the
magnitude of $m_f$. On the other hand, for $v\to 0$, one can see the
equivalence in
other terms as well, but only for $m_f=0$, since the fermion mass is
also proportional to $v$. Examples of such equivalence can be seen in
the calculation of $W^+W^-$ pair production from a lepton-antilepton
pair~\cite{DKW88}.

\paragraph*{Acknowledgements~:} I am grateful to D.~A. Dicus
for showing me his calculations \cite{Dic**} of the process
$\gamma\gamma\to W^+W^-$ using circularly polarized photons and for
calculating the amplitude for $\gamma Z\to W^+W^-$. I also thank
 B. Dutta and S. Nandi for taking time to explain the results of
their papers on  $W^+W^- \to ZZ$ \cite{DuNa94,DuNa94b}. The research
was supported by the Department of Energy of the USA.


\begin{thebibliography}{WW}

\bibitem{CLT74} J.~M. Cornwall, D.~N. Leven, G. Tiktopoulos: Phys. Rev.
D10  (1974) 1145.

\bibitem{LQT77}
B.~W. Lee, C. Quigg, H.~B. Thacker: Phys. Rev. D16  (1977) 1519.

\bibitem{ChGa85}
M. Chanowitz, M.~K. Gaillard: Nucl. Phys.  B261 (1985) 379.

\bibitem{Wil88} S.~S.~D. Willenbrock: Ann. Phys. 186 (1988) 15.

\bibitem{BaSc90} J. Bagger, C. Schmidt: Phys. Rev. D41  (1990) 264.

\bibitem{Vel90} H. Veltman: Phys. Rev. D41 (1990) 2294.

\bibitem{DuNa94} B. Dutta, S. Nandi: ``Test of goldstone boson
equivalence theorem'', Oklahoma State University Preprint 279
(undated), to appear in Mod. Phys. Lett.

\bibitem{DKW88} D. A. Dicus, K.~J. Kallianpur, S.~S.~D. Willenbrock:
Phys. Lett. B200 (1988) 187.

\bibitem{Kal88} K.~J. Kallianpur: Phys. Lett. B215 (1988) 392.

\bibitem{BCHP90} V. Barger, K. Cheung, T. Han, R.~J.~N. Phillips:
Phys. Rev. D42 (1990) 3052.

\bibitem{BeLl87} C. Bento, C.~H. Llewellyn Smith: Nucl. Phys. B289
(1987) 36.

\bibitem{DuNa94b} B. Dutta, S. Nandi: ``Testing goldstone boson
equivalence theorem in hadron colliders'', Oklahoma State University
Preprint 284 (December 1993), to appear in Phys. Rev. D.

\bibitem{ApCa75} T. Appelquist, J. Carrazone: Phys. Rev. D11 (1975) 2856.

\bibitem{BoJi92} E. E. Boos, G.~V. Jikia: Phys. Lett. B275 (1992) 164.

\bibitem{Dic**} D. A. Dicus: private communication.

	\end{thebibliography}
\end{document}